\documentclass[a4paper,twoside,reqno]{bjp-preprint-2307.04678}
\usepackage{graphicx}
\usepackage{cite}
\usepackage{amssymb,amsmath,amscd,amsthm}
\usepackage{times}

\usepackage[bookmarks=false]{hyperref}
\hypersetup{%
    colorlinks=true,        
    linkcolor=blue,          
    citecolor=blue,         
    urlcolor=blue           
    }

\newcommand{\half}{{\textstyle \frac{1}{2}}} 

\usepackage{geometry}
\allowdisplaybreaks

\renewcommand{\thefootnote}{\fnsymbol{footnote}}  

\begin{document}

\vspace*{-5mm}

\title{New Type of Traversable Wormhole$^{\,\S}$}

\addtocounter{footnote}{4}  
\footnotetext{\;Invited paper at
Bahamas Advanced Study Institute \& Conferences (BASIC),\hfill
February 8--14,\\ \vspace*{0.5mm}
\hspace*{-1.25mm}
2023,  Stella Maris Resort, Long Island, Bahamas.
Contribution to appear in the Proceedings, which\\ \vspace*{0.5mm}
\hspace*{-1.mm}
will be published in Bulg. J. Phys.
%
%
}  

\runningheads{Frans R. Klinkhamer}{New Type of Traversable Wormhole}

\begin{start}{%
\author{Frans R. Klinkhamer}{1}

\address{Institute for Theoretical Physics,
 Karlsruhe Institute of Technology (KIT),
 76128 Karlsruhe, Germany}{1}


}

\begin{Abstract}
We review a new traversable-wormhole solution of the gravitational field
equation of general relativity without exotic matter.
Instead of having exotic matter to keep the wormhole throat open,
the solution relies on a
3-dimensional ``spacetime defect,'' which is
characterized by a locally vanishing metric determinant.
We also discuss the corresponding multiple-vacuum-defect-wormhole solution
and possible experimental signatures from a ``gas''
of vacuum-defect wormholes. Multiple vacuum-defect wormholes appear
to allow for backward time travel.
\end{Abstract}

\vspace*{-2mm}

\begin{KEY}
general relativity, wormhole solution
\end{KEY}
\end{start}

\section{Introduction}
\label{sec:Introduction}

\begin{figure}[t] 
\vspace*{0mm}
\centerline{\includegraphics[width=0.75\textwidth]
{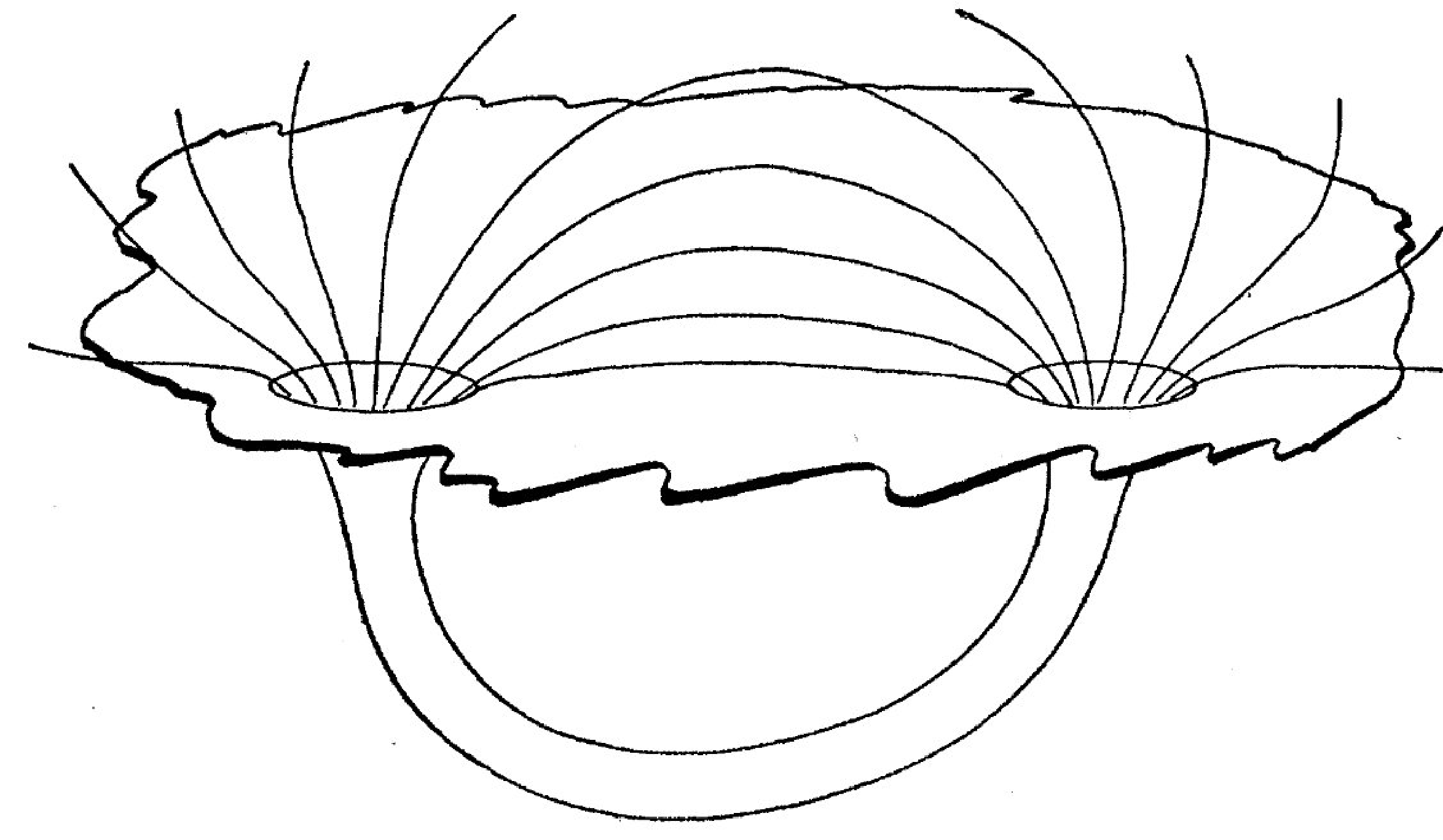}}
\vspace*{-0mm}
\caption{First sketch of a wormhole in the modern
physics literature from Wheeler's 1955 paper.
[Image credit: J.A.~Wheeler, ``Geons,''
Phys. Rev. \textbf{97}, 511 (1955);
DOI: https://doi.org/10.1103/PhysRev.48.73]}
\label{fig:Wheeler-1955-good-screenshot}
\end{figure}

Here is an ultra-brief history of theoretical ideas on
wormholes in five crucial dates:\vspace*{-1mm}
\begin{itemize}
\item 1915,
Einstein's general relativity is established
with spacetime acting as a dynamical quantity
(a summary is published
one year later~\cite{Einstein1916}).\vspace*{-1mm}
\item 1935,
the metric of the Einstein-Rosen ``bridge''
connecting different parts of spacetime is
presented~\cite{EinsteinRosen1935},
but a traveller cannot go across,
as a singularity will be encountered first.\vspace*{-1mm}
\item 1955,
Wheeler~\cite{Wheeler1955} considers a multiply-connected 3-space
and gives a sketch of a ``tunnel'' or ``handle'' in his Figure~7
(here, reproduced as Figure~\ref{fig:Wheeler-1955-good-screenshot});
in the following years, he also discusses a spacetime foam teeming
with (Euclidean) wormholes of Planckian length scales,
$l_{P} \equiv \sqrt{\hbar\, G/c^3} \sim 10^{-35}\,\text{m}$
(a later review paper appears in 1968~\cite{Wheeler1968}).\vspace*{-1mm}
\item 1973,
two papers~\cite{Ellis1973,Bronnikov1973} are published
with a metric suitable for a traversable wormhole.\vspace*{-1mm}
\item 1988,
Morris and Thorne~\cite{MorrisThorne1988} show
that certain wormhole solutions of the Einstein equation
may be traversable, but at a high price: \emph{exotic matter}.\vspace*{-1mm}
\end{itemize}

In this contribution, we discuss a recently discovered
wormhole solution~\cite{Klinkhamer2023a,Klinkhamer2023b},
where the price for traversability
is paid not by the matter but by the spacetime structure.
This new wormhole solution
is based on earlier work~\cite{Klinkhamer2019-regBB},
as will be explained later on.

For completeness, we should perhaps mention that the metric of the
Einstein-Rosen bridge does
\emph{not} solve the vacuum gravitational field equation,
as noted somewhat obliquely in the original paper~\cite{EinsteinRosen1935}.
See, e.g., Ref.~\cite{Guendelman-etal-2009} for further discussion.

\section{Exotic-Matter Wormhole}
\label{sec:Exotic-matter-wormhole}

Morris and Thorne (MT) considered a simple metric
in  Box~2 of their 1988 paper~\cite{MorrisThorne1988}.
Specifically, they discuss the following special case
of a more general metric (setting $c=1$):%
\begin{eqnarray}
\label{eq:EBMT-special}
&&ds^{2}\,\Big|^\text{(EBMT-special)}
\equiv
g_{\mu\nu}(x)\, dx^\mu\,dx^\nu \,\Big|^\text{(EBMT-special)}
\nonumber\\[1mm]
&&=
- dt^{2} + dl^{2}
+ \left(b_{0}^{2} + l^{2}\right)\,
  \Big[ d\theta^{2} + \sin^{2}\theta\, d\phi^{2} \Big]\,,
\end{eqnarray}
with a nonzero real constant $b_{0}$ (taken to be positive, for definiteness).
The coordinates $t$ and $l$ in (\ref{eq:EBMT-special})
range over $(-\infty,\,\infty)$, while the coordinates
$\theta \in [0,\,\pi]$ and $\phi \in [0,\,2\pi)$
are the standard spherical polar coordinates [strictly speaking,
we should use \emph{two} coordinate patches
for the 2-sphere, for example, by stereographic projections from the
North Pole and the South Pole].

Earlier discussions of this type of metric
have appeared in independent papers by Ellis~\cite{Ellis1973}
and Bronnikov~\cite{Bronnikov1973}.
Hence, we have added  ``EB'' to the suffix of
(\ref{eq:EBMT-special}).

\begin{figure}[t]
\vspace*{0mm}
\centerline{\includegraphics[width=0.75\textwidth]
{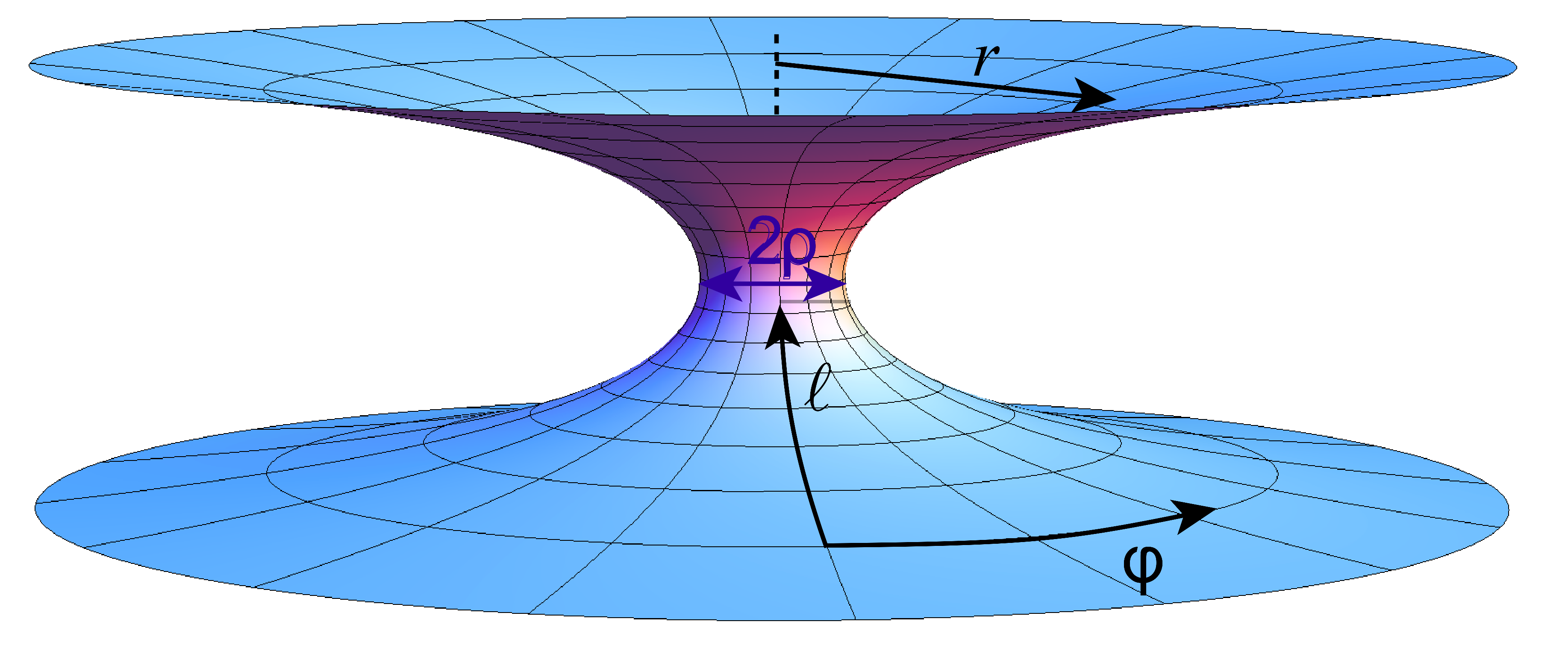}}
\caption{Embedding diagram of the wormhole spacetime with metric
\eqref{eq:EBMT-special} for
$t=\text{const}$ and $\theta=\pi/2$, with temporary definitions
$2 \rho \equiv 2 b_{0}>0$
and $r \equiv \sqrt{b_{0}^{2} + l^{2}}\,$.
[Image credit: O.~James, E.~von Tunzelmann, P.~Franklin, and K.S.~Thorne,
``Visualizing Interstellar's Wormhole,''
arXiv:1502.03809; DOI: https://doi.org/10.48550/arXiv.1502.03809]}
\label{fig:Ellis-wormhole}
\vspace*{-1mm}
\end{figure}

The resulting Ricci and Kretschmann curvature scalars are
\begin{subequations}\label{eq:EBMT-special-R-K}
\begin{eqnarray}
R\,\Big|^\text{(EBMT-special)} &=&
-2\;\frac{b_{0}^{2}}
         {\left(b_{0}^{2} + l^{2}\right)^{2}}\,,
\\[0mm]
K\,\Big|^\text{(EBMT-special)} &=&
12\;\frac{\left(b_{0}^{2}\right)^{2}}
         {\left(b_{0}^{2} + l^{2}\right)^{4}}\,,
\end{eqnarray}
\end{subequations}
both of which are seen to vanish as $l \to \pm\infty$.
Indeed, two distinct flat Minkowski spacetimes are
approached for $l \to \pm\infty$; see Figure~\ref{fig:Ellis-wormhole}.

The EBMT wormhole is traversable,
as shown by items (d) and (e) of Box~2 in Morris and Thorne's
paper~\cite{MorrisThorne1988} and
by Figure~6 in Ellis' paper~\cite{Ellis1973}.
Incidentally, the wormhole of
Figure~\ref{fig:Ellis-wormhole} is of the \emph{inter}-universe type,
connecting two \emph{distinct} asymptotically-flat spaces,
whereas the wormhole of
Figure~\ref{fig:Wheeler-1955-good-screenshot}
is an \emph{intra}-universe wormhole,
connecting to a \emph{single} asymptotically-flat space.
Figure~\ref{fig:Wheeler-1955-good-screenshot} also
illustrates two further properties of a wormhole, namely,
the wormhole ``mouths'' (essentially the boundaries between
the wormhole tube and the near-flat surrounding space)
and the wormhole ``throat'' (the wormhole tube proper
between the wormhole mouths).

The crucial question, however,  concerns the \emph{dynamics}:
can this EBMT wormhole metric be a solution of the Einstein equation?
Morris and Thorne's brilliant idea~\cite{MorrisThorne1988}
was to use an \emph{engineering approach}:
fix the desired specifications and see what it takes.

With the metric (\ref{eq:EBMT-special})
for a traversable wormhole, the Einstein equation,
\begin{equation}
\label{eq:Einstein equation}
R_{\mu\nu} - \half\, g_{\mu\nu}\,R = 8\pi G\:T_{\mu\nu}\,,
\end{equation}
requires these
components of the energy-momentum tensor~\cite{MorrisThorne1988}:
\begin{subequations}\label{eq:T-components-EBMT-special}
\begin{eqnarray}
\label{eq:T-tt-component-EBMT-special}
T^{\,t}_{\;\;\,t}\,\Big|^\text{(EBMT-special)}
&=& \frac{1}{8\pi G}\;
\frac{b_{0}^{2}}{\left(b_{0}^{2} + l^{2}\right)^{2}}\,,
\\[1mm]
T^{\,l}_{\;\;\,l}\,\Big|^\text{(EBMT-special)}
&=& -\frac{1}{8\pi G}\;
\frac{b_{0}^{2}}{\left(b_{0}^{2} + l^{2}\right)^{2}}\,,
\\[1mm]
T^{\,\theta}_{\;\;\,\theta}\,\Big|^\text{(EBMT-special)}
&=& \frac{1}{8\pi G}\;
\frac{b_{0}^{2}}{\left(b_{0}^{2} + l^{2}\right)^{2}}\,,
\\[1mm]
T^{\,\phi }_{\;\;\,\phi}\,\Big|^\text{(EBMT-special)}
&=& \frac{1}{8\pi G}\;
\frac{b_{0}^{2}}{\left(b_{0}^{2} + l^{2}\right)^{2}}\,,
\end{eqnarray}
\end{subequations}
with all other components vanishing.
As the energy density is given
by $\rho = T^{\,tt} =-T^{\,t}_{\;\;\,t}$,
we have $\rho< 0$ from (\ref{eq:T-tt-component-EBMT-special}),
which certainly corresponds to nonstandard matter.
Moreover, we obtain the following inequality
for the radial null vector $\overline{k}^{\,\mu}=(1,\, 1,\,  0,\,0)$:
\begin{equation}
\label{eq:EBMT-special-NEC-radial}
T^{\,\mu}_{\;\;\,\nu}\,
\overline{k}_{\mu}\,\overline{k}^{\,\nu}\,\Big|^\text{(EBMT-special)}
= \frac{1}{8\pi G}\;
\frac{b_{0}^{2}}{\left(b_{0}^{2} + l^{2}\right)^{2}}\;\big[-1-1\big] < 0\,,
\end{equation}
which corresponds to a violation of the Null-Energy-Condition (NEC).

As stressed by Morris and Thorne~\cite{MorrisThorne1988},
the problem is that
it is not clear if the needed exotic matter really exists.

\section{New Wormhole}
\label{sec:New wormhole}

\subsection{Special Ansatz}
\label{subsec:Special-Ansatz}

We now propose a somewhat different metric~\cite{Klinkhamer2023a}:
\begin{eqnarray}
\label{eq:K-special}
&&ds^{2}\,\Big|^\text{(K-special)}
\equiv
g_{\mu\nu}(x)\, dx^\mu\,dx^\nu \,\Big|^\text{(K-special)}
\nonumber\\[1mm]
&&=
- dt^{2} + \frac{\xi^{2}}{\xi^{2}+\lambda^{2}}\;d\xi^{2}
+ \left(b_{0}^{2} + \xi^{2}\right)\,
  \Big[ d\theta^{2} + \sin^{2}\theta\, d\phi^{2} \Big]\,,
\end{eqnarray}
with nonzero real constants $\lambda$ and $b_{0}$
(both taken to be positive, for definiteness)
and coordinates $t$ and $\xi$ ranging over $(-\infty,\,\infty)$.
The resulting Ricci and Kretschmann curvature scalars are
\begin{subequations}\label{eq:K-special-R-K}
\begin{eqnarray}
R\,\Big|^\text{(K-special)} &=&
-2\;\frac{b_{0}^{2}-\lambda^{2}}
         {\left(b_{0}^{2} + \xi^{2}\right)^{2}}\,,
\\[1mm]
K\,\Big|^\text{(K-special)} &=&
12\;\frac{\left(b_{0}^{2}-\lambda^{2}\right)^{2}}
         {\left(b_{0}^{2} + \xi^{2}\right)^{4}}\,,
\end{eqnarray}
\end{subequations}
both of which are finite, smooth, and vanishing as $\xi \to \pm\infty$.

\renewcommand{\thefootnote}{\arabic{footnote}}  
\addtocounter{footnote}{1}      

The metric $g_{\mu\nu}(x)$ from (\ref{eq:K-special})
is \emph{degenerate} with a vanishing determinant
$g(x) \equiv \det[g_{\mu\nu}(x)]$ at $\xi=0$.\footnote{
The metric $g_{\mu\nu}(x)$ from (\ref{eq:EBMT-special}) is
nondegenerate, as its determinant $g(x)$ vanishes nowhere, provided
two appropriate coordinate patches are used for the 2-sphere.}
In physical terms, this 3-dimensional hypersurface at $\xi=0$
corresponds to a ``spacetime defect.''
The terminology is by analogy with crystallographic defects
in an atomic crystal
(these crystallographic defects are typically formed during
a rapid crystallization process).

The new wormhole metric (\ref{eq:K-special})
did not fall out of the sky but
is a direct follow-up of earlier work on a particular
``time defect'' that regularizes
the big bang~\cite{Klinkhamer2019-regBB}.
That particular paper~\cite{Klinkhamer2019-regBB}
contains further references to previous work
on this type of ``spacetime defect.''

We now use Morris and Thorne's engineering approach.
The Einstein equation \eqref{eq:Einstein equation}
for this new metric then requires
the following nonzero energy-momentum-tensor components:
\begin{subequations}\label{eq:T-UPDOWNcomponents-degenmetric-special}
\begin{eqnarray}
\label{eq:T-UPtDOWNt-component-degenmetric-special}
T^{\,t}_{\;\;\,t}\,\Big|^\text{(K-special)}
&=& \frac{1}{8\pi G}\;
\frac{b_{0}^{2}-\lambda^{2}}{\left(b_{0}^{2}+\xi^{2}\right)^{2}}\,,
\\[1mm]
T^{\,\xi}_{\;\;\,\xi}\,\Big|^\text{(K-special)}
&=& -\frac{1}{8\pi G}\;
\frac{b_{0}^{2}-\lambda^{2}}{\left(b_{0}^{2}+\xi^{2}\right)^{2}}\,,
\\[1mm]
T^{\,\theta}_{\;\;\,\theta}\,\Big|^\text{(K-special)}
&=& \frac{1}{8\pi G}\;
\frac{b_{0}^{2}-\lambda^{2}}{\left(b_{0}^{2}+\xi^{2}\right)^{2}}\,,
\\[1mm]
T^{\,\phi }_{\;\;\,\phi}\,\Big|^\text{(K-special)}
&=& \frac{1}{8\pi G}\;
\frac{b_{0}^{2}-\lambda^{2}}{\left(b_{0}^{2}+\xi^{2}\right)^{2}}\,.
\end{eqnarray}
\end{subequations}
There is a subtle point here, namely that
the Einstein equation \eqref{eq:Einstein equation}
at $\xi=0$ is defined
by continuous extension from its limit $\xi \to 0$
(see Ref.~\cite{Klinkhamer2023a} for further discussion).

Compared to the earlier result (\ref{eq:T-components-EBMT-special}),
we see that the previous factors
$b_{0}^{2}$ in the numerators
have been replaced by new factors $(b_{0}^{2}-\lambda^{2})$
in (\ref{eq:T-UPDOWNcomponents-degenmetric-special}).
Starting from $\lambda^{2}=0^{+}$,
these new numerator factors then change sign as
$\lambda^{2}$ increases above $b_{0}^{2}$  and
we no longer require exotic matter.
Indeed, we have from (\ref{eq:T-UPtDOWNt-component-degenmetric-special})
that $\rho =-T^{\,t}_{\;\;\,t} > 0$ for $\lambda^{2} > b_{0}^{2}\,$.
Moreover, we readily obtain, for any null vector $k^{\,\mu}$
and parameters $\lambda^{2} \geq b_{0}^{2}\,$, the inequality
\begin{equation}
\label{eq:EBMT-special-NEC}
T^{\,\mu}_{\;\;\,\nu}\,k_{\mu}\,k^{\,\nu}\,
\Big|^\text{(K-special)}_{\lambda^{2} \geq b_{0}^{2}}
\geq 0\,,
\end{equation}
which verifies the NEC.
Hence, the new wormhole with degenerate metric (\ref{eq:K-special})
for $\lambda^{2} \geq b_{0}^{2}$ does not require exotic matter.

In fact, the special case $\lambda^{2}=b_{0}^{2}$
of the metric (\ref{eq:K-special})
has  the energy-momentum tensor vanishing altogether,
\begin{equation}
\label{eq:T-UPmuUPnu-component-degenmetric-special}
T^{\,\mu}_{\;\;\,\nu}\,\Big|^\text{(K-special)}_{\lambda^{2}=b_{0}^{2}}
= 0\,,
\end{equation}
and so do the curvature scalars (\ref{eq:K-special-R-K}).
In other words, we have
an \emph{exact} wormhole-type solution of the vacuum Einstein equation.
The corresponding spacetime
is flat but different from Minkowski spacetime. How can that be?
The short answer is by having different orientability
(see Sec.~\ref{subsec:Topology-orientability}).

\subsection{Auxiliary coordinates}
\label{subsec:Auxiliary-coordinates}

If, in the metric (\ref{eq:K-special}),
we change the quasi-radial $\xi$ coordinate to
\begin{equation}
\label{eq:ltilde}
\widetilde{l} = \xi\;\sqrt{1+\lambda^{2}/\xi^{2}}
 \in (-\infty,\,-\lambda] \cup [\lambda,\,\infty)\,,
\end{equation}
we get a metric
\begin{equation}
\label{eq:ltilde-metric}
ds^{2}+=
- dt^{2} + d\widetilde{l}^{\;2}
+ \big(b_{0}^{2} + \widetilde{l}^{\;2} - \lambda^{2}\big)\,
  \Big[ d\theta^{2} + \sin^{2}\theta\, d\phi^{2} \Big]\,.
\end{equation}
This last metric is similar to the one from (\ref{eq:EBMT-special}).

But the coordinate transformation $\xi \to \widetilde{l}$
is \emph{not} a diffeomorphism, as the transformation is
discontinuous, jumping from $-\lambda$ to $+\lambda$ at $\xi=0$.
Moreover, the coordinate $\widetilde{l}$ is unsatisfactory
for a proper description of the whole spacetime manifold:
for fixed values of $\{t,\,\theta,\,\phi\}$,
\emph{both} coordinates $\widetilde{l}=-\lambda$ and $\widetilde{l}=\lambda$
correspond to a \emph{single} point of the manifold
(of course, there \emph{is} the single coordinate $\xi=0$).

Still, we can get a useful picture of the spacetime,
as will be discussed in Sec.~\ref{subsec:Topology-orientability}.
Note that the metric \eqref{eq:ltilde-metric}
for $b_{0}^{2}=\lambda^{2}$ has the standard flat form
in terms of spherical coordinates, but the effective squared radius
$\widetilde{l}^{\;2}$ now has a reduced range, $[\lambda^{2},\, \infty)$
for $\lambda^{2} > 0$.

\subsection{Topology and spatial orientability}
\label{subsec:Topology-orientability}

With the coordinates $\big\{\widetilde{l},\,\theta ,\, \phi\big\}$
in the metric (\ref{eq:ltilde-metric}) for general $\lambda>0$
and $b_{0}>0$, we introduce the following two sets of
Cartesian coordinates
[one for the ``upper'' (+) world with $\widetilde{l}>\lambda$
and another for the ``lower''  (-) world with $\widetilde{l}<-\lambda$
]:
\begin{subequations}\label{eq:Cartesian-coordinates}
\begin{eqnarray}
\label{eq:Cartesian-coordinates-plus}
\hspace*{-6mm}
\left\{
\begin{array}{c}
  Z_{+} \\
  Y_{+} \\
  X_{+}
\end{array}
 \right\}
&=& \widetilde{l}\;
\left\{
\begin{array}{l}
  \phantom{Q}\hspace*{-3mm}\cos\theta\\
  \phantom{Q}\hspace*{-3mm}\sin\theta\,\sin\phi \\
  \phantom{Q}\hspace*{-3mm}\sin\theta\,\cos\phi
\end{array}
 \right\}\,, \;\text{for}\;\; \widetilde{l}\geq \lambda>0\,,
\\[1mm]
\label{eq:Cartesian-coordinates-minus}
\hspace*{-6mm}
\left\{
\begin{array}{c}
  Z_{-} \\
  Y_{-} \\
  X_{-}
\end{array}
 \right\}
&=& \widetilde{l}\;
\left\{
\begin{array}{l}
  \phantom{Q}\hspace*{-3mm}\cos\theta\\
  \phantom{Q}\hspace*{-3mm}\sin\theta\,\sin\phi \\
  \phantom{Q}\hspace*{-3mm}\sin\theta\,\cos\phi
\end{array}
 \right\}\,, \;\text{for}\;\; \widetilde{l}\leq -\lambda<0\,,
\\[1mm]
\label{eq:Cartesian-coordinates-antipodal}
\hspace*{-6mm}
 \left\{Z_{+},\,  Y_{+},\,  X_{+}\right\}\,\Big|_{\,\widetilde{l}=+\lambda}
&\;\stackrel{\wedge}{=}\;&
 \left\{Z_{-},\,  Y_{-},\,  X_{-}\right\}\,\Big|_{\,\widetilde{l}=-\lambda}\,,
\end{eqnarray}
\end{subequations}
where the last relation implements the identification of
``antipodal'' points on the two \mbox{2-spheres} $S^{\,2}_{\pm}$
with $\widetilde{l}=\pm\lambda$.

\begin{figure}[t]   
\vspace*{-0mm}
\centerline{\includegraphics[width=0.60000\textwidth]
{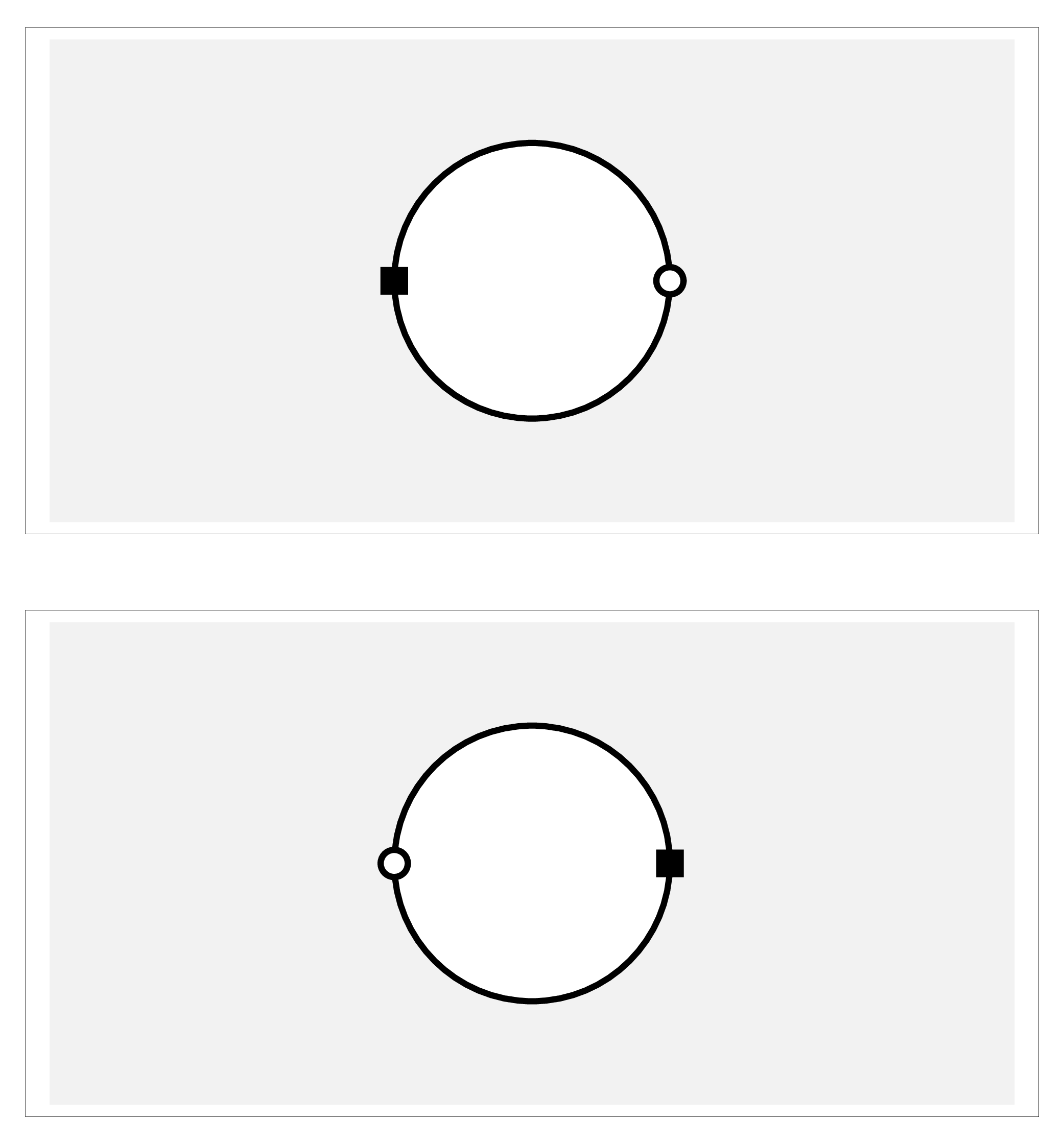}}
\vspace*{-0mm}
\caption{Topology of the spacetime with a single defect wormhole,
at $Z_{\pm}=0$ and an arbitrary fixed time $t$.
The ``upper'' ($+$) world is shown by
the top panel in this figure and the ``lower'' ($-$) world
by the bottom panel.
A single defect wormhole connects the two worlds.
Here, the two wormhole mouths are
shown as two heavy circles with ``antipodal'' spacetime points identified
(two distinct points on the wormhole mouths
are marked by two different symbols, the filled square and the small
circle). The  ``throat'' of this particular wormhole has zero length,
different from the wormholes sketched in
Figures~\ref{fig:Wheeler-1955-good-screenshot}
and \ref{fig:Ellis-wormhole}.}
\label{fig:Topology-single-defect-wormhole-points}
\vspace*{00mm}
\end{figure}

The spatial topology of our degenerate-wormhole spacetime (\ref{eq:K-special})
is that of two copies of $\mathbb{R}^{3}$
with the interior of two balls removed
and ``antipodal'' identification (\ref{eq:Cartesian-coordinates-antipodal})
of their two surfaces (see Figure~\ref{fig:Topology-single-defect-wormhole-points}).
Note that the two coordinates sets
$\left\{Z_{\pm},\,  Y_{\pm},\,  X_{\pm}\right\}$
from (\ref{eq:Cartesian-coordinates-plus})
and (\ref{eq:Cartesian-coordinates-minus})
have \emph{different orientation}.

It can be verified that the
defect-wormhole spacetime from (\ref{eq:K-special}) and
(\ref{eq:Cartesian-coordinates}) is simply connected
(all loops in 3-space can be contracted to a point).
The defect-wormhole topology is, therefore, different
from that of the original exotic-matter EBMT wormhole,
which is multiply connected, having
noncontractible loops in space (for example, a
loop in the upper world encircling once the wormhole mouth).

\subsection{Radial geodesics}
\label{subsec:Radial geodesics}

From the vacuum-wormhole metric (\ref{eq:K-special})
with $b_{0}^{2}=\lambda^{2}$,
we can get explicitly the radial geodesics $\xi(t)$
passing through the wormhole throat at $\xi=0$:
\begin{subequations}\label{eq:radial-geodesics}
\begin{eqnarray}
\hspace*{-5mm}
\theta(t)\,\Big|^\text{(K-special)}_\text{vacuum\;sol\,;\;rad-geod} &=& \pi/2\,,
\\[1mm]
\hspace*{-5mm}
\phi(t)\,\Big|^\text{(K-special)}_\text{vacuum\;sol\,;\;rad-geod}  &=& 0\,,
\\[1mm]
\hspace*{-5mm}
\xi(t)\,\Big|^\text{(K-special)}_\text{vacuum\;sol\,;\;rad-geod}
&=&
\begin{cases}
 \pm\,\sqrt{(B\,t)^{2}+2\,B\,\lambda\,t} \,,   &  \;\;\text{for}\;\; t \geq 0 \,,
 \\[1mm]
 \mp\,\sqrt{(B\,t)^{2}-2\,B\,\lambda\,t} \,,   &  \;\;\text{for}\;\; t \leq 0 \,,
\end{cases}
\end{eqnarray}
\end{subequations}
with different signs (upper or lower)
in front of the square roots for motion in opposite directions
and a dimensionless constant $B \in (0,\,1]$.

\begin{figure}[t]
\vspace*{-0mm}
\centerline{\includegraphics[width=0.80\textwidth]
{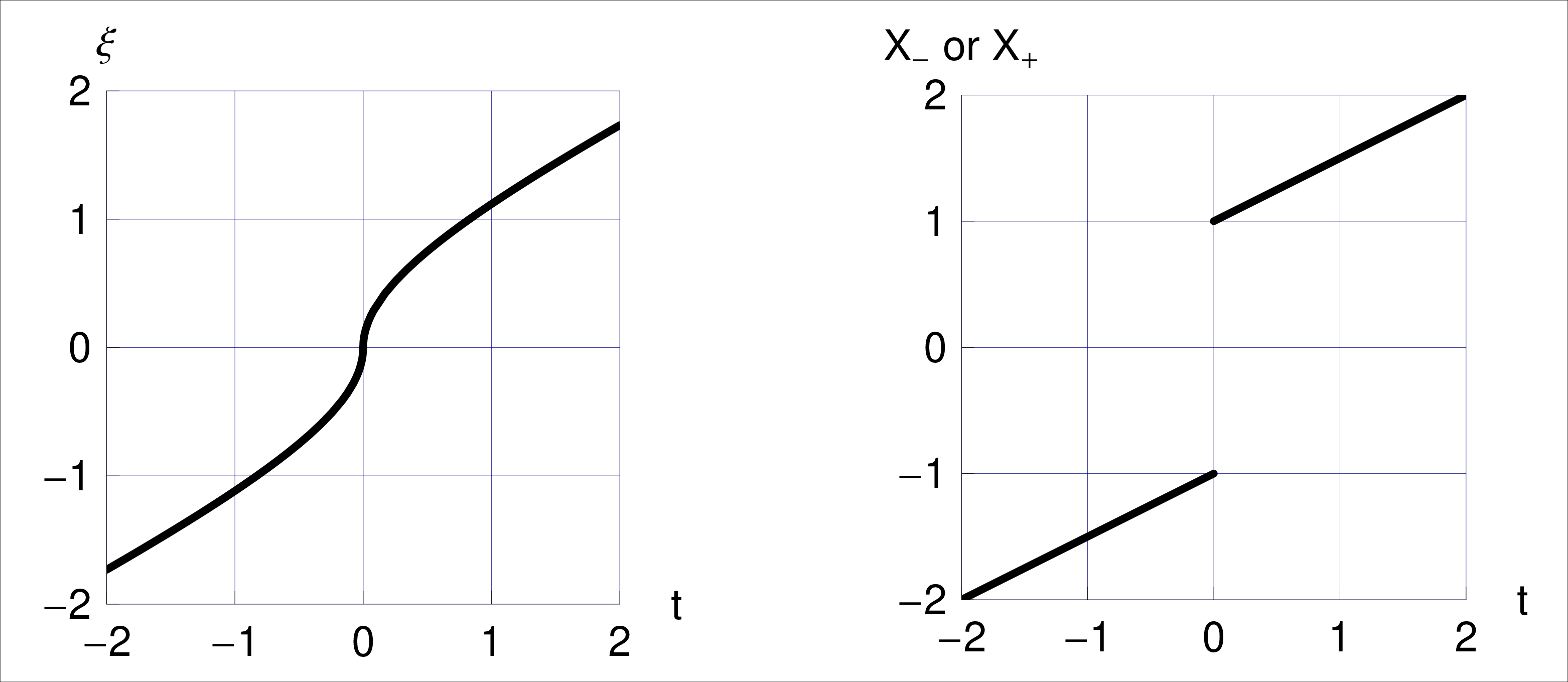}}
\caption{Radial geodesic
(\ref{eq:radial-geodesics}) and (\ref{eq:radial-geodesic-Cartesian-coord})
with $\lambda=1$ and $B=1/2$.}
\label{fig:plotradialgeodesicvacwormhole}
\end{figure}

The radial geodesic (\ref{eq:radial-geodesics})
with the upper signs has the following
trajectory in terms of the
Cartesian coordinates (\ref{eq:Cartesian-coordinates}):
\begin{subequations}\label{eq:radial-geodesic-Cartesian-coord}
\begin{eqnarray}
\hspace*{-5mm}
Z_{\pm}(t)\,\Big|^\text{(K-special)}_\text{vacuum\;sol\,;\;rad-geod}
&=&  0\,,\qquad\qquad\;\; \text{for}\;\; t \in (-\infty,\,\infty)\,,
\\[1mm]
\hspace*{-5mm}
Y_{\pm}(t)\,\Big|^\text{(K-special)}_\text{vacuum\;sol\,;\;rad-geod}
&=&  0\,,\qquad\qquad\;\; \text{for}\;\; t \in (-\infty,\,\infty)\,,
\\[1mm]
\hspace*{-5mm}
X_{-}(t)\,\Big|^\text{(K-special)}_\text{vacuum\;sol\,;\;rad-geod}
&=&
 -\lambda+B\,t\,,\;\;\;\;\text{for}\;\; t \leq 0 \,,
\\[1mm]
\hspace*{-5mm}
X_{+}(t)\,\Big|^\text{(K-special)}_\text{vacuum\;sol\,;\;rad-geod}
&=& +\lambda+B\,t\,,\;\;\;\;\text{for}\;\; t \ge 0 \,,
\end{eqnarray}
\end{subequations}
with $X_{-}=-\lambda$ and $X_{+}=+\lambda$ identified at $t=0$.
The curves in the  $(t,\,X_{-})$ and $(t,\,X_{+})$ planes
have two parallel straight-line segments, shifted at $t=0$,
with equal constant positive slope $B\leq 1$
(velocity magnitude in units with $c=1$);
see Figure~\ref{fig:plotradialgeodesicvacwormhole}.

This equal Minkowski-spacetime velocity before and after the defect crossing
is the main argument for using the ``antipodal''
identification in (\ref{eq:Cartesian-coordinates}),
rather than a ``direct'' identification
on the two 2-spheres $S^{\,2}_{\pm}$,
which would correspond to replacing the prefactors $\widetilde{l}$ in
(\ref{eq:Cartesian-coordinates-plus})
and (\ref{eq:Cartesian-coordinates-minus}) by $|\,\widetilde{l}\,|$
and would have a unique spatial orientation (but apparently nonsmooth
motion along defect-crossing geodesics).
A further argument in favor of the ``antipodal'' identification,
at least for the vacuum case, is given by
demanding a smooth solution of
the first-order equations of general relativity~\cite{Klinkhamer2023a}.

\subsection{General Ansatz}
\label{subsec:General-Ansatz}

The special degenerate metric (\ref{eq:K-special})
can be generalized as follows~\cite{Klinkhamer2023a}:
\begin{eqnarray}
\label{eq:K}
&&ds^{2}\,\Big|^\text{(K-general)}
\equiv
g_{\mu\nu}(x)\, dx^\mu\,dx^\nu \,\Big|^\text{(K-general)}
\nonumber\\[1mm]
&&=
- e^{2\,\widetilde{\phi}(\xi)}\;dt^{2}
+ \frac{\xi^{2}}{\xi^{2}+\lambda^{2}}\;d\xi^{2}
+ \widetilde{r}^{\;2}(\xi)\,
  \Big[ d\theta^{2} + \sin^{2}\theta\, d\phi^{2}  \Big]\,,
\end{eqnarray}
with a positive length scale $\lambda$ and
real functions $\widetilde{\phi}(\xi)$ and $\widetilde{r}(\xi)$.
Again, the coordinates $t$ and $\xi$
range over $(-\infty,\,\infty)$, while
$\theta \in [0,\,\pi]$ and $\phi \in [0,\,2\pi)$
are the standard spherical polar coordinates
[as mentioned before, we should really use two
appropriate coordinate patches for the 2-sphere].
If we assume that
$\widetilde{\phi}(\xi)$ remains finite everywhere
and that $\widetilde{r}(\xi)$ is positive with
$\widetilde{r}(\xi) \sim |\xi |$ for $\xi \to \pm\infty$,
then the spacetime from
(\ref{eq:K}) corresponds to a wormhole.

If the function $\widetilde{r}(\xi)$ has a global minimum
at the value $b_{0}>0$ for $\xi=0$
and if the redshift function $\widetilde{\phi}(\xi)$
is essentially constant near $\xi=0$,
then we can expect to obtain nontrivial dynamics
for $\lambda^{2}$ values of the order of $b_{0}^{2}$ or larger.
As mentioned in Ref.~\cite{Klinkhamer2023a},
we have once used power series in $\xi^{2}$ for
$\widetilde{\phi}(\xi)$ and $\widetilde{r}^{\;2}(\xi)$
and have found energy-momentum components without singular
behavior at $\xi=0$. Further work remains to be done.

\section{Single Vacuum-Defect Wormhole Revisited}
\label{sec:Single-vacuum-defect-wormhole-revisited}

We already have an \emph{exact} wormhole-type solution of the
vacuum Einstein gravitational field equation,
as mentioned in the last paragraph of Sec.~\ref{subsec:Special-Ansatz}
for the special-case metric.
Here, we review the vacuum-defect-wormhole solution
(abbreviated ``vac-def-WH-sol'') in terms of the
tetrad and show that it provides
a smooth  solution~\cite{Klinkhamer2023a}
of the first-order equations of general relativity.
The relevant fields are the tetrad $e^{a}_{\mu}(x)$,
which builds the metric tensor $g_{\mu\nu}(x) \equiv
e^{a}_{\phantom{z}\mu}(x)\,e^{b}_{\phantom{z}\nu}(x)\,\eta_{ab}$,
and the Lorentz connection $\omega^{\phantom{z}a}_{\mu\phantom{z}b}(x)$.

With the differential-form formalism in the notation
of Ref.~\cite{EguchiGilkeyHanson1980}
and the definition of the curvature 2-form
\begin{equation}
\label{eq:curvature-2-form}
R^{\,a}_{\,\phantom{z}b} \equiv \text{d}\omega^{\,a}_{\,\phantom{z}b}+
\omega^{a}_{\phantom{z}c}  \wedge \omega^{c}_{\phantom{z}b}\,,
\end{equation}
the first-order vacuum equations of general relativity
are  explicitly~\cite{Horowitz1991}%
\begin{subequations}\label{eq:first-order-eqs}
\begin{eqnarray}\label{eq:first-order-eqs-no-torsion}
e^{\,[\,a} \wedge D\, e^{\,b\,]} &=& 0\,,
\\[1mm]
\label{eq:first-order-eqs-Ricci-flat}
e^{\,b} \wedge R^{\,cd}\,\epsilon_{abcd} &=& 0 \,,
\end{eqnarray}
\end{subequations}
where the square brackets around Lorentz indices $a$ and $b$
denote antisymmetrization.
Furthermore, $\epsilon_{abcd}$ is the completely antisymmetric symbol
and the covariant derivative is defined by
$D\, e^{b} \equiv \text{d}e^{b}+\omega^{\,b}_{\,\phantom{z}c}\wedge e^{c}$.
The first equation of \eqref{eq:first-order-eqs}
corresponds to the no-torsion condition
and the second to the Ricci-flatness equation.

For the vacuum-defect wormhole, we make the following
\textit{Ans\"{a}tze} for the dual basis (defined in terms of the tetrad
by $e^{a} \equiv e^{a}_{\phantom{z}\mu}\,\text{d}x^{\mu}$)\,:
\begin{subequations}\label{eq:vacuum-wormhole-tetrad}
\begin{eqnarray}
\label{eq:vacuum-wormhole-tetrad-a-is-0}
e^{0}\,\Big|_\text{vac-def-WH-sol}
&=& \text{d}t\,,
\\[1mm]
\label{eq:vacuum-wormhole-tetrad-a-is-1}
e^{1}\,\Big|_\text{vac-def-WH-sol}
&=&  \frac{\xi}{\sqrt{b^{2} + \xi^{2}}}\; \text{d}\xi\,,
\\[1mm]
e^{2}\,\Big|_\text{vac-def-WH-sol}
&=&  \sqrt{b^{2} + \xi^{2}}\; \text{d}\theta\,,
\\[1mm]
e^{3}\,\Big|_\text{vac-def-WH-sol}
&=& \sqrt{b^{2} + \xi^{2}}\;\sin\theta  \;  \text{d}\phi\,,
\end{eqnarray}
\end{subequations}
and for the nonzero components of the connection 1-form:
\begin{eqnarray}
\label{eq:vacuum-wormhole-connection}
\left\{
\omega^{2}_{\phantom{z}1},\,
\omega^{3}_{\phantom{z}1},\,
\omega^{3}_{\phantom{z}2}
\right\}\,\Big|_\text{vac-def-WH-sol}
&=&
\left\{
-\omega^{1}_{\phantom{z}2},\,
-\omega^{1}_{\phantom{z}3},\,
-\omega^{2}_{\phantom{z}3}
\right\}\,\Big|_\text{vac-def-WH-sol}
\nonumber\\[1mm]
&=&
\left\{
\text{d}\theta,\,
\sin\theta  \;  \text{d}\phi,\,
\cos\theta  \;  \text{d}\phi
\right\}\,.
\end{eqnarray}
This particular tetrad and connection are perfectly smooth
(in particular, at $\xi=0$) and give a vanishing curvature 2-form:
\begin{equation}
\label{eq:vacuum-wormhole-curvature}
R^{\,a}_{\phantom{z}b}\,\Big|_\text{vac-def-WH-sol}
=0\,,
\end{equation}
which corresponds to having the Riemann tensor
$\mathcal{R}_{\kappa\lambda\mu\nu}(x)=0$
in the standard coordinate formulation
(the calligraphic symbol marks the difference with
the curvature 2-form $R^{\,a}_{\phantom{z}b}$).

We can, now, verify that the two
first-order vacuum equations in \eqref{eq:first-order-eqs}
are solved.
The first (no-torsion) equation is solved, as
it has been used to construct the connection components
\eqref{eq:vacuum-wormhole-connection}
from the tetrad \textit{Ansatz} \eqref{eq:vacuum-wormhole-tetrad}.
The second (Ricci) equation is trivially solved
by \eqref{eq:vacuum-wormhole-curvature}.

\section{Multiple Vacuum-Defect Wormholes}
\label{sec:Multiple-vacuum-defect-wormholes}

The construction of the multiple-vacuum-defect-wormhole solution
(abbreviated ``multiple-vac-def-WH-sol'') is straightforward.
Here, we only give a brief heuristic discussion
and point to Ref.~\cite{Klinkhamer2023b} for technical details.

We start by defining
the embedding space $M_\text{embed}$,
which corresponds to the union of two copies of
Euclidean 3-space $E_{3}$, one copy being labeled by `$+$'
(the ``upper'' world) and the other by `$-$' (the ``lower'' world).
In each of these 3-spaces, we have standard Cartesian coordinates,
\begin{subequations}
\begin{eqnarray}
\label{eq:embedding-space-def}
M_\text{embed} &=&
E_{3}^{\,(+)} \cup \; E_{3}^{\,(-)}
\\[2mm]
\label{eq:embedding-space-coordinates}
E_{3}^{\,(\pm)} &:&
\left(X_{\pm},\,  Y_{\pm},\,  Z_{\pm}\right) \;\in\; \mathbb{R}^{3}\,.
\end{eqnarray}
\end{subequations}
The multiple-vacuum-defect-wormhole solutions only cover
part of $M_\text{embed}$.

Next, we define $4N$ solution parameters,
\begin{subequations}\label{eq:multiple-vac-def-WH-sol-parameters}
\begin{eqnarray}
b_{n} &>&  0,\, \;\;\;
\quad
\text{for}\;\;n \in \{1,\, 2,\, \ldots\,,\,  N\}\,,
\\[2mm]
\left(\widehat{X}_{n},\, \widehat{Y}_{n},\,\widehat{Z}_{n}\right)
&\in& \mathbb{R}^{3}\,,
\quad
\text{for}\;\;n \in \{1,\, 2,\,  \ldots\,,\,  N\}\,,
\end{eqnarray}
where $b_{n}$ is the length scale
(throat circumference divided by $2\pi$)
of the $n$-th wormhole
and $(\widehat{X}_{n},\, \widehat{Y}_{n},\,\widehat{Z}_{n})$
are the corresponding center coordinates in the embedding space
(equal for lower and upper world). These parameters must be
such that balls $B_{n}$ around the $n$-th center with radius $b_{n}$
do not touch each other,
\begin{eqnarray}
B_{1} \cap B_{2}  \cap \,\ldots\, \cap B_{N} &=& \emptyset\,.
\end{eqnarray}
\end{subequations}
The typical size of the wormhole mouths
will be denoted by $\overline{b}$ and
the typical separation of the wormhole mouths
by $\overline{l}$, which is best defined by the average wormhole
number density $\overline{n}\equiv 1/\overline{l}^{\,3}$.

Finally, we perform surgery on the embedding space by removing the
interiors of the balls $B_{n}$ and identifying
``antipodal'' points on the respective 2-sphere boundaries,
similar to the identification in
\eqref {eq:Cartesian-coordinates-antipodal}
with $\lambda$ replaced by $b_{n}$.

\begin{figure}[t]   
\vspace*{-0mm}
\centerline{\includegraphics[width=0.60000\textwidth]
{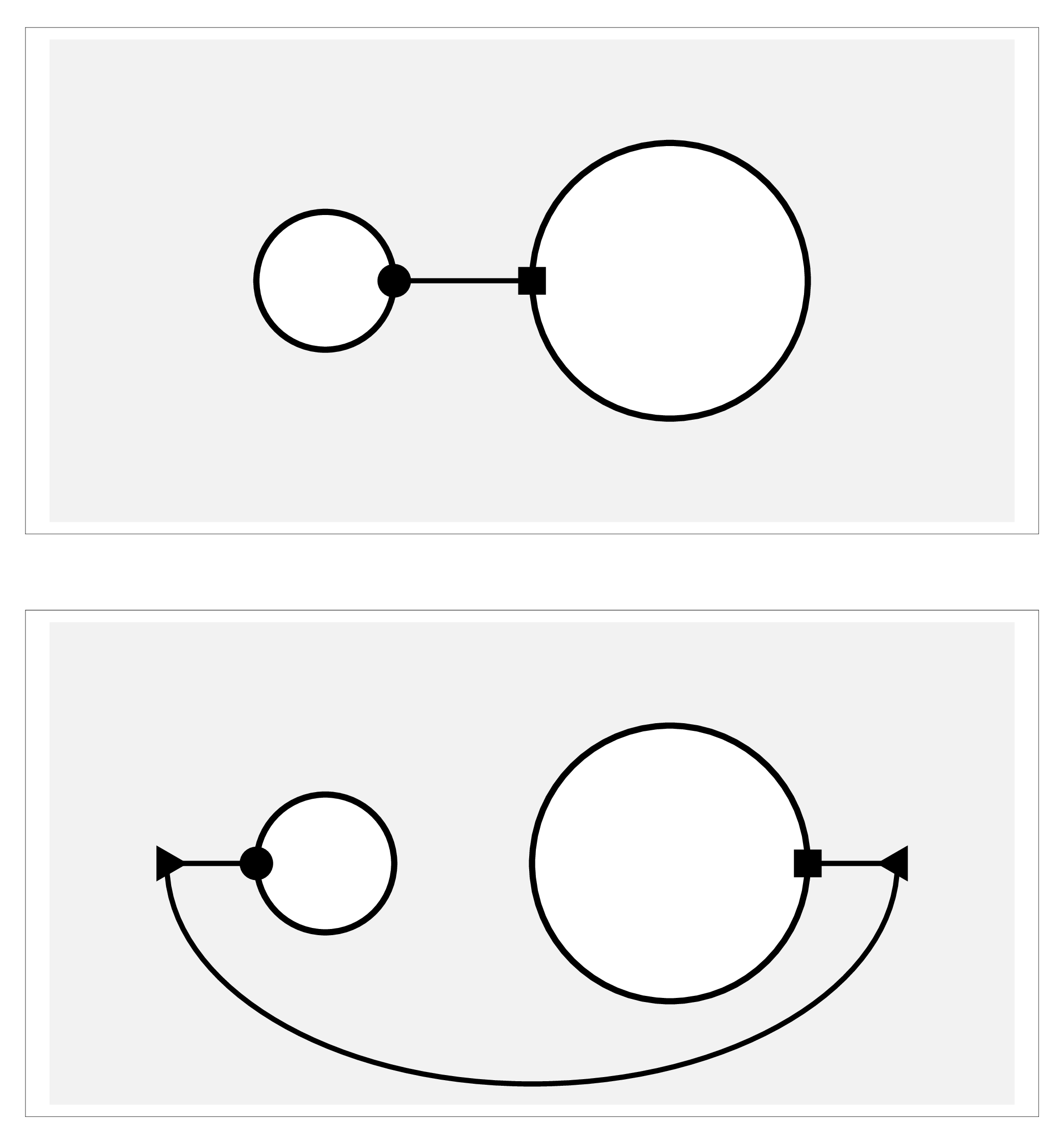}}
\vspace*{-0mm}
\caption{Sketch of the $N=2$
multiple-vacuum-defect-wormhole spacetime,
at $Z_{\pm}=0$ and an arbitrary fixed time $t$.
The top panel in this figure corresponds to the
``upper'' world (label `$+$')
and the bottom panel to the ``lower'' world (label `$-$').
The two worlds are connected by vacuum-defect wormholes.
Here, the wormhole mouths
are shown as heavy circles (having radius $b_{1}$ for the left
wormhole and $b_{2}$ for the right one)
with ``antipodal'' spacetime points identified
(two distinct points are marked by two different symbols,
the heavy dot and the filled square).
To travel between two particular points
of the ``lower'' world (shown by two different triangles),
there is a long way staying in the ``lower'' world
and a short way though the two wormhole throats with a short
passage in the ``upper'' world.
}
\label{fig:N-is-2-multiple-wormhole-points}
\vspace*{00mm}
\end{figure}

This completes the basic construction of the coordinates and
a sketch appears in
Figure~\ref{fig:N-is-2-multiple-wormhole-points} for $N=2$.
We now turn to the metric of the
multiple-vacuum-defect-wormhole solution.

Between the wormhole mouths, there is the standard
flat metric in either 3-space (labelled by $\pm$),
\begin{equation}
\label{eq:outside-metric}
ds^{2}\,\Big|_\text{multiple-vac-def-WH-sol}^{\text{(outside\;WH-mouths)}}
=
- dt^{2} + \left(d X_{\pm} \right)^{2}
+ \left(d Y_{\pm} \right)^{2}
+ \left(d Z_{\pm} \right)^{2}
\,,
\end{equation}
for the spatial coordinates \eqref{eq:embedding-space-coordinates}
of the embedding space.

For the metric at or near the wormhole mouths,
we have essentially the single-defect-wormhole
metric \eqref{eq:K-special} for $\lambda=b_{0}=b_{n}$.
Specifically, the nearby metric for
the particular wormhole mouth with label $\overline{n}$ is given by
\begin{eqnarray}
\label{eq:near-WH-mouth-metric}
&&ds^{2}\,\Big|_\text{multiple-vac-def-WH-sol}^{\text{(near WH-mouth $\overline{n}$)}}
\nonumber\\[2mm]
&&
=
- dt^{2}
+ \frac{\xi_{\,\overline{n}}^{2}}{b_{\,\overline{n}}^{2}
+ \xi_{\,\overline{n}}^{2}}\;d\xi_{\,\overline{n}}^{2}
+ \left(b_{\,\overline{n}}^{2} + \xi_{\,\overline{n}}^{2}\right)\,
  \Big[ d\theta_{\,\overline{n}}^{2}
        + \sin^{2}\theta_{\,\overline{n}}\, d\phi_{\,\overline{n}}^{2} \Big]\,,
\end{eqnarray}
with coordinates
$\xi_{\,\overline{n}} \in [-\Delta_{\,\overline{n}},\,\Delta_{\,\overline{n}}]$
for a positive infinitesimal $\Delta_{\,\overline{n}}\,$,
$\theta_{\,\overline{n}} \in [0,\,\pi]$, and
$\phi_{\,\overline{n}} \in [0,\,2\pi)$.

\section{Phenomenological Aspects}
\label{sec:Phenomenological-aspects}

\subsection{Preliminaries}
\label{subsec:Preliminaries}

If there is a ``gas'' of randomly positioned
static vacuum-defect wormholes, we would like to measure,
or at least bound, the typical wormhole size $\overline{b}$
and the typical wormhole separation
$\overline{l} \equiv (\overline{n})^{-1/3}$, defined in terms of
the average number density $\overline{n}$.
We have shown in Ref.~\cite{Klinkhamer2023b}
that, different from the case of
similar space defects in a single spacetime,
vacuum-Cherenkov and scattering bounds
on $\overline{b}$ and $\overline{l}$ appear to be rather
ineffectual. More successful are imaging bounds,
to which we will now turn.

\subsection{Imaging bound from an electron microscope}
\label{subsec:Imaging-bound-from-electron-microscope}

\begin{figure}[t]   
\vspace*{-0mm}
\centerline{\includegraphics[width=0.60000\textwidth]
{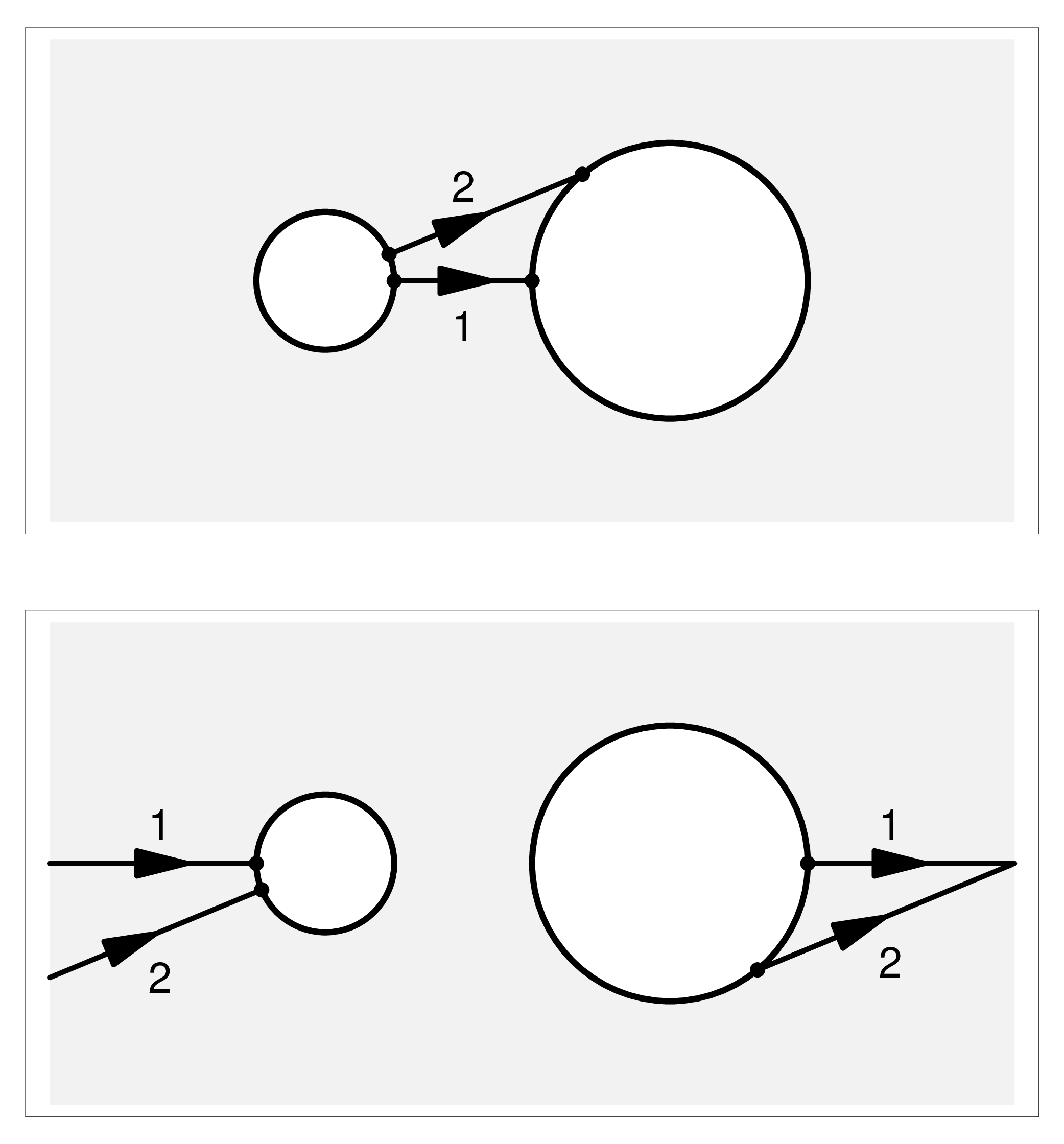}}
\vspace*{-0mm}
\caption{Two light rays in the
$N=2$ multiple-vacuum-defect-wormhole spacetime from
Figure~\ref{fig:N-is-2-multiple-wormhole-points},
with a flat metric given by \eqref{eq:outside-metric}
between the wormhole mouths.
The light rays start in the ``lower'' world on the left
and the small dots on the wormhole mouths are purely indicative.
Both rays re-emerge in the ``lower'' world after a passage
in the ``upper'' world. The special ray marked ``$1$''
(connecting the two fictitious wormhole centers)
continues straight on, but the other ray marked ``$2$''
is parallel shifted (here, by a translation to the right with a
magnitude equal to twice the wormhole length scale $b_{2}$).
}
\label{fig:N-is-2-multiple-wormhole-light-rays}
\vspace*{00mm}
\end{figure}

For light travelling a distance $L$, there are
$N \sim L/\overline{l}$ encounters with the wormhole mouths.
In each encounter, the ray is typically parallel shifted
by an amount of order $\pm b_{n}$, which is illustrated in
Figure~\ref{fig:N-is-2-multiple-wormhole-light-rays}
for a single pair of wormholes ($N=2$).
As explained in Sec.~IV B of Ref.~\cite{Klinkhamer2023b},
the built-up parallel shift of the light ray
results from a random-walk process
and its order of magnitude is given by
\begin{equation}
\label{eq:Delta-x}
[\Delta x]_\text{random-walk} \sim \sqrt{N}\,\overline{b}
\sim \left(\,\overline{b}/\overline{l}\,\right)\;
\sqrt{L\,\overline{l}}\,.
\end{equation}
In order to get a clear image of a 2D sample with substructure
$[\delta x]_\text{sample}$, we must have negligible random-walk
parallel shifts of the light ray,
\begin{equation}
\label{eq:clear-image-condition}
[\Delta x]_\text{random-walk} \lesssim [\delta x]_\text{sample}\,,
\end{equation}
where the inequality should perhaps be stronger
(with `$\lesssim$' replaced by `$\ll$'),
but we prefer to remain conservative.

Instead of light rays we can also consider electron beams
in a transmission electron microscope (TEM);
see, e.g., Ref.~\cite{Spence2017} for some background.
A special-purpose TEM would have condenser lenses designed
to produce more or less parallel rays after the rays have passed
through a 2D sample with sub\r{a}ngstr\"{o}m structure.
Furthermore, these parallel rays just after the 2D sample must travel
freely for about $L \sim 10\,\text{cm}$ (possibly broadened
by vacuum-defect-wormhole effects) before they enter the rest of the microscope.
If this experimental setup could indeed be realized
and if clear images would be obtained, then
the following bound would result~\cite{Klinkhamer2023b}:
\begin{equation}
\label{eq:upper-bound-electron-microscope}
\left(\,\overline{b}/\overline{l}\,\right)\;\overline{b}\;
\Big|^\text{(special-TEM)}
\lesssim
\frac{\left([\delta x]_\text{sample}\right)^{2}}{L}
=
10^{-21}\,\text{m}\;
\left(\frac{[\delta x]_\text{sample}}{0.1\,\text{\r{A}}}\right)^{2}\,
\left(\frac{0.1\,\text{m}}{L}\right)\,,
\end{equation}
where \eqref{eq:Delta-x}
and \eqref{eq:clear-image-condition} have been used.

Bound \eqref{eq:upper-bound-electron-microscope},
for the $[\delta x]_\text{sample}$ and $L$ values stated,
is perfectly consistent with having a gas of
Planck-scale vacuum-defect-wormholes,
$\,\overline{b} \sim \overline{l} \sim 10^{-35}\,\text{m}$.

\subsection{Backward time travel}
\label{subsec:backward-time-travel}

It was pointed out in
Morris--Thorne's landmark paper~\cite{MorrisThorne1988} that
traversable exotic-matter wormholes appear to
allow for backward time travel.
An alternative procedure for getting a time machine
was suggested by Frolov and Novikov~\cite{FrolovNovikov1990}.
These last authors propose to insert
a large point mass near one of the mouths of an
intra-universe traversable exotic-matter wormhole, so that a clock near
that mouth runs slower than a clock near
the other mouth and a time machine appears after a sufficiently long time.

A similar time machine can be constructed
by use of multiple vacuum-defect wormholes
and a single localized mass~\cite{Klinkhamer2023b}.
Essentially, there are two steps.

In the first step, we observe that a \emph{pair}
of vacuum-defect wormholes
can act as a single \emph{intra}-universe wormhole.
Consider, namely, a point (left triangle) in the lower world near
the heavy dot of the bottom panel in
Figure~\ref{fig:N-is-2-multiple-wormhole-points}.
While staying in the lower world, it is possible
to travel along a long path
to a point (right triangle) near the filled square on the right
(see the ellipse segment between the two triangles in the
bottom panel of Figure~\ref{fig:N-is-2-multiple-wormhole-points}).
Alternatively, it is possible,  in the lower world, to enter
the left wormhole  at the heavy dot,
to emerge in the upper world at the heavy dot,
to travel in the upper world along a short straight line
towards the filled square,
to enter the right wormhole at the filled square,
and, finally, to re-emerge in the lower world at the filled square.

In the second step,
we position, for an identical wormhole pair
(Figure~\ref{fig:N-is-2-multiple-wormhole-points}
but now with open symbols, the open square and the small circle),
a static point mass in the lower world just to the
right of the open square in the bottom panel of the modified
Figure~\ref{fig:N-is-2-multiple-wormhole-points}. Now, a
lower-world clock near the open square runs slower
(gravitational redshift)
than a lower-world clock near the small circle.
Recall that the gravitational-redshift effect
follows directly from the equivalence principle;
see, e.g., Sec.~7.4 of Ref.~\cite{MisnerThorneWheeler2017}.

\begin{figure}[t]   
\centerline{\includegraphics[width=0.80\textwidth]
{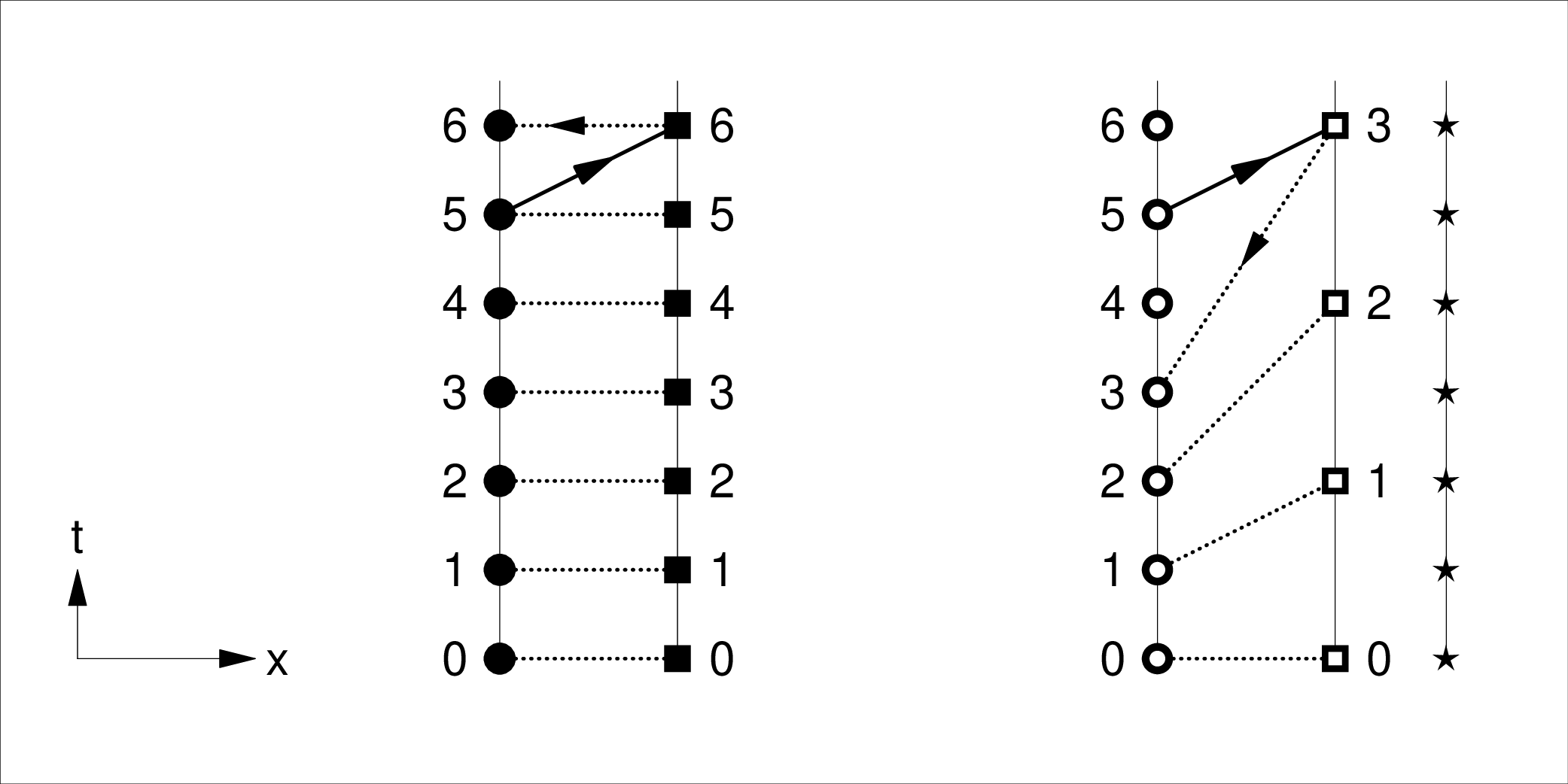}}
\caption{Spacetime-diagram sketches (in rescaled units)
of four identical lower-world clocks
near the four wormhole mouths of two identical vacuum-defect-wormhole pairs.
The figure
explains the appearance of a time machine for the vacuum-defect-wormhole
pair on the right with an added lower-world static point mass
(star) near the right wormhole mouth (open square);
see Sec.~\ref{subsec:backward-time-travel} for a detailed description.}
\label{fig:N-is-2-time-machine}
\end{figure}

After these two steps, the time machine appears
automatically and an illustration is given by
Figure~\ref{fig:N-is-2-time-machine}. As said,
take two identical vacuum-defect-wormhole pairs
(the first pair with the heavy dot and the
filled square in the bottom panel of
Figure~\ref{fig:N-is-2-multiple-wormhole-points} and the
second pair from the same figure but now with open symbols).
Introduce four identical lower-world clocks,
at rest near the four wormhole mouths.
Next, consider two paths
starting at the left triangle in the bottom panel of
Figure~\ref{fig:N-is-2-multiple-wormhole-points}
and ending at the right triangle in the bottom panel:
the long path (ellipse segment) that remains entirely
in the lower world
and the short path that first enters the left wormhole,
runs in the upper world towards the right wormhole,
enters the right wormhole, and
re-emerges in the lower world. In order to streamline
the further discussion, we move the left triangle
in Figure~\ref{fig:N-is-2-multiple-wormhole-points} close
to the heavy dot and the right triangle close to
the filled square. Also, we push the
two wormholes together, so that the straight path
in the upper world is negligibly short compared to the
long path in the lower world.

Without adding a point mass
(left wormhole pair in Figure~\ref{fig:N-is-2-time-machine}),
both clocks run at the same rate and the ticks are shown for
$t_\text{left}=t_\text{right}=0,\,1,\,\ldots \,,\,6$.
An ultrafast explorer starts out at local time
$t_\text{left}=5$ from a point near the left wormhole mouth
(heavy dot) on the long path in the lower world
(full line with arrow in the spacetime diagram on the left)
to reach a point near the right wormhole mouth (filled square)
and then passes quickly through the two wormhole throats
(dotted line with arrow in the spacetime diagram on the left) back to
the starting point (heavy dot) at local time $t_\text{left}=6$.

With a static point mass added (shown by a star for the right
wormhole pair in Figure~\ref{fig:N-is-2-time-machine}), the clock
near the right wormhole mouth (open square)
runs slower than the clock
near the left wormhole mouth (small circle).
Again, an explorer starts out at local time
$t_\text{left}=5$ from a point near the left wormhole mouth
(small circle) on the long path in the lower world
(full line with arrow in the spacetime diagram on the right)
to reach a point near the right wormhole mouth (open square)
and then passes quickly through the two wormhole throats
(dotted line with arrow in the spacetime diagram  on the right)
back to
the starting point (small circle) at local time $t_\text{left}=3$.
In other words, the explorer has travelled back in time
($t_\text{left,\,return}=3$ being less than $t_\text{left,\,start}=5$)
and effectively a time machine is operating.
In fact, the time machine starts working for
departure times $t_\text{left} \geq 1$
(the same explorer leaving at local time
$t_\text{left}=1$ returns at local time $t_\text{left}=1$).
For earlier departure times $t_\text{left}<1$, the
return is at a later time $t_\text{left}$. For example, if the
same traveller leaves at $t_\text{left}=0$, then the
return will be at $t_\text{left}=1/2$.

This simple example shows that, in principle, a time machine can
appear if there exists one pair of vacuum-defect wormholes
and a single point mass which can be freely positioned
in either of the two worlds.
Whether or not this time machine without exotic matter is stable
(classically and quantum mechanically) remains to be seen.
Obviously, backward time travel is hard to swallow and we refer
to, e.g., Ref.~\cite{Visser1996} for further discussion.

\section{Final Remarks}
\label{sec:Final-remarks}

The vacuum-defect-wormhole solution (\ref{eq:vacuum-wormhole-tetrad})
and \eqref{eq:vacuum-wormhole-connection}
has a length scale $\lambda$ as a free parameter.
If there is  a preferred value $\overline{\lambda}$ in
the actual Universe, then that value
can only come from a theory beyond general relativity.

One such theory could be nonperturbative superstring theory
in a matrix-model realization. We have explicitly considered
the
Ishibashi--Kawai--Kitazawa--Tsu\-chi\-ya
(IKKT) matrix model~\cite{IKKT-1997},
also known as IIB matrix model~\cite{Aoki-etal-review-1999}.
That matrix model could indeed give rise to an emergent spacetime
with or without spacetime
defects~\cite{Klinkhamer2021-master,Klinkhamer2022-corfu}.
If defects do appear,
then the typical length scale $\overline{\lambda}$ of a remnant
vacuum-wormhole defect would be related
to the IIB-matrix-model length scale $\ell$.
The Planck length $l_{P} \equiv G^{1/2}$ would
also be related to $\ell$. Combined, this
suggests that the typical vacuum-defect-wormhole length scale
$\overline{\lambda}$ might be of the order
of the Planck length $l_{P} \sim 10^{-35}\,\text{m}$.

At this moment, we do not have experimental bounds
which rule out a ``gas'' of Planck-scale vacuum-defect
wormholes (Sec.~\ref{subsec:Imaging-bound-from-electron-microscope}).
For such Planck-scale wormholes, the appearance of backward time travel
(Sec.~\ref{subsec:backward-time-travel}) is perhaps acceptable.

If it is now possible to ``harvest''
such Planck-scale vacuum-defect wormholes and to ``fatten'' them
by adding a finite amount of normal matter (as suggested by the
results of App.~B in Ref.~\cite{Klinkhamer2023a}),
then we may have to live up to
the possibility of having time machines
for human-scale robots (and possibly humans themselves).
But all this is admittedly still very far off in the future,
if at all realistic.

\section*{Note Added}

An interesting extension of the vacuum-defect-wormhole solution
has been suggested recently by Wang~\cite{Wang2023}.
His suggestion is to add a
Schwarzschild-type factor $\big[1-2M/\sqrt{b^{2} + \xi^{2}}\,\big]^{1/2}$
to the dual-basis
component \eqref{eq:vacuum-wormhole-tetrad-a-is-0}
and the inverse factor to the dual-basis
component \eqref{eq:vacuum-wormhole-tetrad-a-is-1},
where the parameter $M < |b|/2$ can be positive or negative.
The corresponding
spin-connection 1-form $\omega^{a}_{\phantom{z}b}$
follows directly from the
no-torsion condition \eqref{eq:first-order-eqs-no-torsion}.
These extended \textit{Ans\"{a}tze} for the tetrad and
connection then solve the
Ricci-flatness equation \eqref{eq:first-order-eqs-Ricci-flat},
as shown in App.~B of Ref.~\cite{Wang2023}.

This mathematical result is more or less obvious,
but the physical interpretation is less clear.
In the original Schwarzschild vacuum metric,
we interpret $M$ as a point
mass sitting at $r=0$, where the vacuum metric no longer holds.
But what is the physics of $M$ in the
extended-vacuum-defect-wormhole spacetime?
There appears to be no place for the matter,
as the manifold is geodesically complete.
One possible interpretation may be that a nonzero $M$
in the 4-dimensional metric describes higher-dimensional effects.

\section*{Acknowledgements}
The author thanks Thomas Curtright, Eduardo Guendelman, and Peter West
for organizing the BASIC2023 meeting and for accepting an online
presentation.

\end{document}